\documentstyle[12pt,twoside,fleqn,espcrc1]{article}

\newcommand{\AmS}{{\protect\the\textfont2
  A\kern-.1667em\lower.5ex\hbox{M}\kern-.125emS}}
\newcommand{\gtrsim}{\:\lower 0.4ex\hbox{$\stackrel{\scriptstyle >}
  {\scriptstyle\sim}$}\:}
\newcommand{\lesssim}{\:\lower 0.4ex\hbox{$\stackrel{\scriptstyle <}
  {\scriptstyle\sim}$}\:}
\newcommand{\sig}{\:\lower 0.6ex\hbox{$\stackrel{\textstyle >}
  {\textstyle\sim}$}\:}
\newcommand{\sil}{\:\lower 0.6ex\hbox{$\stackrel{\textstyle <}
  {\textstyle\sim}$}\:}

\title{Neutrino-nucleus interaction and supernova $r$-process 
       nucleosynthesis}

\author{Yong-Zhong Qian\address{Physics Department, 161-33, 
        California Institute of Technology,\\ 
        Pasadena, CA 91125, USA}\thanks{
	Supported by the David W. Morrisroe Fellowship
	at Caltech.}}

\begin{document}
\maketitle

\begin{abstract}
We discuss various neutrino-nucleus interactions  
in connection with the supernova $r$-process nucleosynthesis, which
possibly occurs in the neutrino-driven wind of a young neutron star. 
These interactions include $\nu_e$ and $\bar\nu_e$
absorptions on free nucleons, $\nu_e$ captures on neutron-rich nuclei,
and neutral-current $\nu_{\mu(\tau)}$ and $\bar\nu_{\mu(\tau)}$
interactions with $\alpha$-particles and neutron-rich nuclei.
We describe how these interactions can affect the $r$-process nucleosynthesis
and discuss the implications of their effects for the physical conditions
leading to a successful supernova $r$-process.
We conclude that a low electron fraction and/or a short
dynamic time scale may be required to give the sufficient
neutron-to-seed ratio for an $r$-process in the neutrino-driven wind.
In the case of a short dynamic time scale, the wind has to be contained
during the $r$-process. Possible mechanisms which can give a low electron
fraction or contain the wind are discussed.
\end{abstract}

\section{INTRODUCTION}

In this paper we discuss the role of neutrino-nucleus interactions in a
recent model of the supernova $r$-process nucleosynthesis. In this model, the
$r$-process occurs in the neutrino-driven wind of a young neutron star
\cite{woohoff}.
The winds leave the neutron star at times $t\gtrsim 1$ s after its creation by
the supernova explosion, and last for $\sim 10$ s, the Kelvin-Helmholtz
cooling time scale of the neutron star. The attractive feature of this
model is the simple characterization of its physical conditions by the
neutron star properties and the supernova neutrino flux \cite{qiawoo}.
The physical conditions relevant for the $r$-process nucleosynthesis
are the electron fraction, the entropy per baryon, the dynamic time
scale, and the mass loss rate in the wind. These conditions are determined
mostly by $\nu_e$ and $\bar\nu_e$ absorptions on the free nucleons in the
wind. However, the presence of significant
neutrino flux also introduces various other interaction processes
between neutrinos and nuclei, which prove to be interesting and important
for the possible $r$-process nucleosynthesis in this model. In Sec. 2,
we describe some salient features of the $r$-process nucleosynthesis
in the neutrino-driven wind. In Sec. 3, we focus on the role of
$\nu_e$ and $\bar\nu_e$
absorptions on free nucleons in determining
the electron fraction in the wind.
In Sec. 4, we investigate the influence of 
the $\alpha$-particles in the wind on setting
the electron fraction and the neutron-to-seed ratio for the $r$-process. In
particular, we describe how neutrino spallations on 
the $\alpha$-particles can alter
the path of the nuclear flow leading to the seed nuclei. 
We discuss the requirement on the conditions in the wind to avoid or
counteract various effects concerning the $\alpha$-particles. In Sec. 5,
we examine the competition between $\nu_e$ captures on neutron-rich
nuclei and nuclear $\beta$-decays during the $r$-process. We discuss
the implications of this competition for the dynamic time scale and
the location of the $r$-process. We also briefly mention the post-processing
of the $r$-process abundance distribution through neutron
emissions by the nuclei highly excited in $\nu_e$ captures and 
neutral-current neutrino interactions.
We summarize our discussions and give conclusions
in Sec. 6.

\section{POSSIBLE $r$-PROCESS IN THE NEUTRINO-DRIVEN WIND}

For the lack of a good astrophysical model, many studies 
characterize the conditions for the $r$-process with
the neutron number density and the temperature at the
freeze-out of the $r$-process 
(see e.g., \cite{kratz}). 
Sometimes a neutron
exposure time is also introduced. The $r$-process abundance distributions
produced under a set of different conditions are then superposed with 
appropriate weights to fit the observed solar $r$-process
abundance distribution. In the recent model of the $r$-process
in the neutrino-driven wind, the important physical parameters are the
electron fraction, $Y_e$, the entropy per baryon, $S$, the dynamic time
scale, $\tau_{\rm dyn}$, and the mass loss rate, $\dot M$, in the wind.
All these parameters are determined mostly by $\nu_e$ and $\bar\nu_e$
absorptions on the free nucleons in the wind at temperatures much higher than
what is relevant for the $r$-process.
Once these parameters are fixed, the adiabatic
expansion of a particular mass element in the wind specifies its temperature,
$T$, and density, $\rho$, as it moves away from the neutron star. It is
convenient to think of the radius of this mass element, $r$, as a time 
evolution parameter. We can relate the time, $t$, to $r$ through $dt=dr/v$,
with $v$ the velocity of the mass element.
For example, in a radiation-dominated wind with
$\tau_{\rm dyn} \sim r/v \sim$ constant, from $\dot M=4\pi r^2\rho v$ and
$S\propto T^3/\rho$, we have $T\propto r^{-1}$, $\rho\propto r^{-3}$, and
$r\propto \exp(t/\tau_{\rm dyn})$ to describe the adiabatic expansion
of the mass element. 
The nucleosynthesis in the mass element
can then be characterized by $Y_e$, $S$, and $\tau_{\rm dyn}$,
with the neutron excess given by $1-2Y_e$, and 
the time evolution of $T$ and $\rho$ specified by
$S$ and $\tau_{\rm dyn}$. In other
words, these three parameters determine the neutron number density
and the temperature at the freeze-out of the $r$-process
and the neutron exposure time
in the neutrino-driven wind. 
The parameters $Y_e$, $S$, $\tau_{\rm dyn}$, and
$\dot M$ in the wind
depend on the supernova neutrino flux. 
The supernova neutrino characteristics evolve on time scales 
$\sim 1\ {\rm s}\gg \tau_{\rm dyn}$, which 
naturally leads to a superposition of the $r$-process abundance
distributions produced under various conditions within a single astrophysical
event. The mass loss rate determines the relative $r$-process contribution 
made with the corresponding $Y_e$, $S$, and $\tau_{\rm dyn}$.
When integrated
over the neutron star Kelvin-Helmholtz cooling time scale, 
it also gives the
total amount of the $r$-process material ejected in each supernova. 
Therefore, in principle, this model could reproduce
the observed solar $r$-process
abundance distribution and account for the galactic $r$-process yields
without fine-tuning any arbitrary parameter \cite{woo}. 

A simplistic picture for the physical processes leading to the $r$-process
in the neutrino-driven wind is as follows. At $T_9\gtrsim10$ ($T_9$ stands for
the temperature in $10^9$ K), the wind material is essentially composed of
free nucleons. The absorptions of $\nu_e$ and $\bar\nu_e$ on free neutrons
and protons, respectively, are the main heating mechanisms to drive
the expansion of the material. In the meantime, these absorption reactions
also govern the evolution of the electron fraction at these temperatures.
At $10\gtrsim T_9\gtrsim5$, 
nuclear statistical equilibrium approximately holds, and
the wind material mostly consists of free nucleons and $\alpha$-particles.
All the physical parameters in the wind have been set by the time the
temperature reaches $T_9\sim5$. In general, the mass loss rate is set
first, followed by the entropy and the dynamic time scale. The electron
fraction is set last \cite{qiawoo}. For $Y_e<0.5$ and $S>100$ per baryon,
an $\alpha$-process takes place at $5\gtrsim T_9\gtrsim3$, 
resulting in seed nuclei
of mass number $A_s\sim100$ \cite{woohoff}. 
At $3\gtrsim T_9\gtrsim1$, the $r$-process
occurs under neutron-rich and possibly $\alpha$-rich conditions.
In the following sections, we discuss the role of neutrino-nucleus
interactions in the various physical processes described above.
As we will see, these discussions complicate the simple picture of the
$r$-process in the neutrino-driven wind. Nevertheless, they also provide
a better understanding of the physical conditions for a successful
supernova $r$-process.

\section{DETERMINATION OF THE ELECTRON FRACTION}

Obviously, the $r$-process only operates under neutron-rich conditions.
In other words, an $r$-process is possible only if the electron fraction
is set at $Y_e<0.5$. At $T_9\gtrsim10$, the following reactions are important
for setting the electron fraction in the neutrino-driven wind:
\begin{eqnarray}
\nu_e+n&\rightleftharpoons&p+e^-,\\
\bar\nu_e+p&\rightleftharpoons&n+e^+.
\end{eqnarray}
To first approximation, the evolution of the electron fraction 
in the wind is determined by 
\begin{equation}
{dY_e\over dt}\equiv v{dY_e\over dr}=\lambda_1-\lambda_2 Y_e,
\end{equation}
where $\lambda_1=\lambda_{\nu_en}+\lambda_{e^+n}$, and $\lambda_2
=\lambda_1+\lambda_{\bar\nu_ep}+\lambda_{e^-p}$. We denote the rates
for the forward reactions in Eqs. (1) and (2) as $\lambda_{\nu_en}$
and $\lambda_{\bar\nu_ep}$, respectively. The rates for the corresponding
reverse reactions are denoted by $\lambda_{e^-p}$ and $\lambda_{e^+n}$,
respectively. It can be shown that as long as the rate $\lambda_2$ is larger 
than the expansion rate of the wind material, the electron fraction 
is approximately given by $Y_e\approx \lambda_1/\lambda_2$ \cite{qiawoo}.
If this still holds when the temperature-sensitive rates $\lambda_{e^-p}$
and $\lambda_{e^+n}$ become negligible compared with the neutrino
absorption rates $\lambda_{\bar\nu_ep}$ and $\lambda_{\nu_en}$, 
the electron fraction depends on the ratio 
$\lambda_{\bar\nu_ep}/\lambda_{\nu_en}$ alone, 
and is given by 
\begin{equation}
Y_e\approx \left(1+{\Phi_{\bar\nu_e}\over\Phi_{\nu_e}}
{\langle E_{\bar\nu_e}^2\rangle-2\Delta \langle E_{\bar\nu_e}\rangle+\Delta^2
\over\langle E_{\nu_e}^2\rangle+2\Delta \langle E_{\nu_e}\rangle+\Delta^2}
\right)^{-1},
\end{equation}
where for example, $\Phi_{\nu_e}$ and $\langle E_{\nu_e}^n\rangle$ are
the $\nu_e$ flux and the $n$th moment of the normalized $\nu_e$ energy
spectrum, respectively. The neutron-proton mass difference, $\Delta=1.293$
MeV, enters Eq. (4) through the dependence of 
the neutrino absorption cross sections on the phase space of the outgoing
leptons \cite{qiawoo}. 

Equation (4) connects the electron fraction in the wind with the 
characteristics of supernova neutrino emission. The neutron-rich
material inside the neutron star gives rise to
different opacities for $\nu_e$ and $\bar\nu_e$
through the forward reactions in Eqs. (1) and (2). As a result,
$\bar\nu_e$ decouple at higher temperatures inside the neutron
star than $\nu_e$, and correspondingly have a harder spectrum.
In terms of the average neutrino energy, $\langle E_\nu\rangle$,
and the effective neutrino energy, $\epsilon_\nu\equiv
\langle E_\nu^2\rangle/\langle E_\nu\rangle$, we have
$\langle E_{\bar\nu_e}\rangle>\langle E_{\nu_e}\rangle$ and
$\epsilon_{\bar\nu_e}>\epsilon_{\nu_e}$. The difference between
the $\bar\nu_e$ and $\nu_e$ spectra increases 
as the neutron star becomes more and more neutron rich with
time. The neutrino flux can be written as $\Phi_\nu\propto
L_\nu/\langle E_\nu\rangle$, with $L_\nu$ the neutrino luminosity.
Numerical supernova neutrino transport calculations show that
$L_{\bar\nu_e}\approx L_{\nu_e}$ for the first 20 s of the neutron
star cooling phase \cite{woo}. During this time, Eq. (4) can
be rewritten as $Y_e\approx (\epsilon_{\nu_e}+2\Delta)/
(\epsilon_{\bar\nu_e}+\epsilon_{\nu_e})$, with
the terms proportional to $\Delta^2$ neglected. Therefore, an $r$-process
can possibly occur only if $\epsilon_{\bar\nu_e}-\epsilon_{\nu_e}
>4\Delta\approx 5.2$ MeV. This corresponds to $t\gtrsim 3$ s after
the supernova explosion (see Fig. 5 in Ref. \cite{qiawoo}). Typically,
we have $\epsilon_{\bar\nu_e}\approx 22$ MeV and 
$\epsilon_{\nu_e}\approx 12$ MeV, which give $Y_e\approx 0.43$.
Of course,
when the $r$-process actually takes place also depends on other
physical parameters in the wind, such as the entropy per baryon.

It is conceivable that the approximate equality of $L_{\bar\nu_e}$
and $L_{\nu_e}$ breaks down when the deleptonization of the neutron
star is nearly complete. This may occur between 20 and 30 s after
the supernova explosion. At these very late times, the approximate equality
of $\Phi_{\bar\nu_e}$ and $\Phi_{\nu_e}$ is likely to hold instead.
According to Eq. (4), the electron fraction in the wind becomes 
even more sensitive to the difference between the $\bar\nu_e$ 
and $\nu_e$ spectra. In this case, very low values of $Y_e$ may be 
obtained. With the same neutrino spectra typical of earlier times,
and assuming $\epsilon_\nu\propto\langle E_\nu\rangle$,
we have $Y_e\approx0.29$. We note that the larger difference between
the $\bar\nu_e$ and $\nu_e$ spectra at later times may give an even
lower $Y_e$.
This could prove extremely helpful when the desirable 
entropy per baryon in the wind is hard to achieve, and/or the
effects considered in the next section become important.  

\section{INFLUENCE OF THE $\alpha$-PARTICLES}

It was shown in Ref. \cite{qiawoo} that the neutrino absorption reactions
in Eqs. (1) and (2) are also responsible for setting the entropy per baryon,
the dynamic time scale, and the mass loss rate in the wind. Typical
entropies are $S\sim 100$ per baryon, and the wind is 
radiation-dominated \cite{qiawoo}.
For entropies of $S\gtrsim 100$ per baryon, 
photo-dissociations of nuclei heavier than $\alpha$-particles are still
strong at $T_9\lesssim10$ due to the significant 
presence of energetic photons on the tail
of the Bose-Einstein distribution. Therefore, $\alpha$-particles become
an important part of the nuclear composition in the wind at $T_9<10$.

In the presence of $\alpha$-particles, Eq. (3) for the evolution of $Y_e$
in the wind is modified to be
\begin{equation}
{dY_e\over dt}=\lambda_1^\prime-\lambda_2 Y_e,
\end{equation}
where $\lambda_1^\prime=(1-X_\alpha)\lambda_1+(X_\alpha/2)\lambda_2$,
with $X_\alpha$ the mass fraction of $\alpha$-particles. In deriving
Eq. (5), we have assumed that free nucleons and $\alpha$-particles
dominate the nuclear composition of the wind. This approximation is
good for $10\gtrsim T_9\gtrsim 5$. For typical dynamic time scales 
in the wind, the rate $\lambda_2$ is still
significant at $T_9\sim 5$. So the final value of $Y_e$ is affected
by the presence of $\alpha$-particles. The effect of the $\alpha$-particles
can be seen from the instantaneous equilibrium value of 
$Y_e=(\lambda_1/\lambda_2)(1-X_\alpha)+X_\alpha/2$
corresponding to Eq. (5). For $\lambda_1/\lambda_2<1/2$,
$\alpha$-particles tend to increase $Y_e$ above the value given by Eq. (4).
This undesirable effect on $Y_e$ for the $r$-process in the wind
was first pointed out in Ref. \cite{full}. 
To avoid this effect, one must have a short dynamic time
scale, for which $Y_e$ freezes out at $T_9\gtrsim10$
when $X_\alpha\approx 0$. In the absence of readily available mechanisms
to achieve such dynamic time scales, we may have to consider 
a low ratio $\lambda_1/\lambda_2$ to counteract the increase in $Y_e$
from the effect of $\alpha$-particles. Without invoking exotic neutrino physics
such as antineutrino oscillations \cite{qia}, we may have to consider
the senario of getting low $Y_e$
at very late times suggested in the previous section.
The large abundance of $\alpha$-particles presents yet another serious effect
on the potential $r$-process nucleosynthesis in the wind, which we
discuss next.
 
In the successful $r$-process calculation of Ref. \cite{woo}, $\alpha$-rich
conditions were obtained with entropies of $S>400$ per baryon in the wind.
We can understand its success by estimating the neutron-to-seed ratio
prior to the $r$-process obtained in the calculation. The seed
nuclei come from the so-called $\alpha$-process, which occurs at
$5\gtrsim T_9\gtrsim 3$. To make heavy seed nuclei from $\alpha$-particles,
the nuclear flow has to pass the bottle-neck of the three-body reactions
$\alpha+\alpha+\alpha\rightarrow$ $^{12}$C and $\alpha+\alpha+n\rightarrow$
$^9$Be. Once the bottle-neck is passed, further $\alpha$-capture reactions
proceed efficiently. By the time the Coulomb barrier stops the charged-particle 
reactions at $T_9\sim3$, the nuclear flow has reached seed nuclei of mass 
number $A_s\sim100$. Because of the inefficiency of 
the three-body reactions, the final
$\alpha$-particle mass fraction, $X_{\alpha,f}$, is still high 
at the end of the $\alpha$-process. From charge conservation, we have
$Y_e\approx X_{\alpha,f}/2+(Z_s/A_s)X_s$, with $Z_s\sim 35$ the charge
of the seed nuclei. At $T_9\sim5$, the wind material mostly consists of
free neutrons and $\alpha$-particles, with the neutron mass fraction
$X_n\approx1-2Y_e$. At $5\gtrsim T_9\gtrsim3$, neutrons are also captured
to make heavy seed nuclei, and $X_n$ decreases somewhat during the
$\alpha$-process. The neutron-to-seed ratio prior to the $r$-process
is approximately given by
\begin{equation}
{n\over s}\approx{X_n\over X_s/A_s}\sil{1-2Y_e\over Y_e-(X_{\alpha,f}/2)}Z_s.
\end{equation}
As we can see, a high $\alpha$-particle mass fraction at the end of the
$\alpha$-process means a large neutron-to-seed ratio, which is required
to produce the $r$-process abundance peaks at mass numbers $A\sim130$ and
195.

However, in the presence of significant neutrino flux, the following
neutral-current neutrino spallation reaction can occur:
\begin{equation}
\nu+\alpha\rightarrow t+p+\nu^\prime.
\end{equation}
Because $\nu_{\mu(\tau)}$ and $\bar\nu_{\mu(\tau)}$ do not have the
charged-current absorption reactions similar to those 
of $\nu_e$ and $\bar\nu_e$ in Eqs. (1) and (2),
they decouple at the highest temperatures inside the neutron star and
have the hardest spectra. They are mainly responsible for the spallation
reactions in Eq. (7). Once a tritium nucleus is produced, the subsequent
$\alpha$-capture reactions $t+\alpha\rightarrow$ $^7$Li and $^7$Li $+\alpha
\rightarrow$ $^{11}$B can bypass the bottle-neck of the three-body reactions
and therefore, expedite the processing of $\alpha$-particles into heavy
seed nuclei. It was found in Ref. \cite{mey} that up to 30 $\alpha$-particles
can disappear as a result of a single spallation reaction. If a significant
fraction of the $\alpha$-particles experience neutrino spallations during
the $\alpha$-process, the reduction of $X_{\alpha,f}$ and the simultaneous
excessive production of seed nuclei will decrease the neutron-to-seed
ratio. In fact, for the conditions employed in the $r$-process calculation
of Ref. \cite{woo}, it was found that the neutron-to-seed ratio becomes
insufficient for effective production of the abundance peak at $A\sim195$
after neutrino spallations on the $\alpha$-particles are included \cite{mey}. 

To avoid significant neutrino spallations on $\alpha$-particles, we need
a short dynamic time scale to reduce the duration of the $\alpha$-process.
Otherwise, we have to require a low $Y_e$, which compensates for a low
$X_{\alpha,f}$, to restore the $r$-process abundance peak at $A\sim195$
in the calculation of Ref. \cite{woo}. On another note, the high entropies
of $S>400$ per baryon in Ref. \cite{woo}
still evades a simple physical explanation. 
Typical entropies in the wind are found to be $S\sim 100$ per baryon
\cite{qiawoo}. For such relatively low entropies, $X_{\alpha,f}$ is
already low even without considering the neutrino spallations, unless
the dynamic time scale in the wind is extremely short, 
$\tau_{\rm dyn}\ll 0.1$ s
\cite{hoff}. Therefore, a short dynamic time scale 
or a low $Y_e$ in the wind is
helpful to the $r$-process in both the high and the low entropy scenarios.

\section{EFFECTS OF $\nu_e$ CAPTURES ON NUCLEI}

The influence of the neutrino flux does not stop at the end of the
$\alpha$-process. In fact, the effects of supernova neutrinos can be
important during and even after the $r$-process. In the standard picture
(see e.g., \cite{kratz}),
the progenitor nuclei on the $r$-process path are in 
$(n,\gamma)\rightleftharpoons(\gamma,n)$ equilibrium during the
$r$-process. The nuclear flow proceeds from one isotopic chain with
charge $Z$ to the next with charge $Z+1$ via nuclear $\beta$-decays.
However, in the presence of significant neutrino flux, the following
reaction can compete with nuclear $\beta$-decays:
\begin{equation}
\nu_e+A(Z,N)\rightarrow A(Z+1,N-1)+e^-,
\end{equation}
where $A(Z,N)$ stands for a nucleus with $Z$ protons, $N$ neutrons,
and mass number $A=Z+N$. It was first pointed out in Ref. \cite{nad}
that $\nu_e$ captures on the progenitor nuclei can potentially speed
up the $r$-process. The $\beta$-decay rate strongly depends on the
parent-daughter ground state mass difference, and can vary by more
than one order of magnitude for different progenitor nuclei. The
slowest $\beta$-decays occur at the so-called waiting-point nuclei
with magic neutron numbers $N=50$, 82, and 126. In fact, the duration
of the $r$-process in the standard picture is essentially controlled by
the $\beta$-decays of these nuclei. Therefore, $\nu_e$ captures on
the waiting-point nuclei are of particular interest. It turns out
that for the same $\nu_e$ luminosity and spectrum,
the $\nu_e$ captures rates are roughly the same
for all the waiting-point nuclei 
\cite{full,qhlv}. If the
duration of the $r$-process is determined by the $\nu_e$ captures 
instead of the $\beta$-decays, the number of $\nu_e$ captures during
the $r$-process must satisfy $\Delta Z>1$. Assuming that the $r$-process
occurs from $T_9\sim 3$ to $T_9\sim 1$ in the neutrino-driven wind,
we have
\begin{equation}
\Delta Z\sim \lambda_{\nu_e}(r_{\rm FO})\int_{r_{\rm FO}/3}^{r_{\rm FO}}
\left({r_{\rm FO}\over r}\right)^2{dr\over v}
\sim 4\lambda_{\nu_e}(r_{\rm FO})\tau_{\rm dyn}>1,
\end{equation}
where $\lambda_{\nu_e}(r_{\rm FO})$ is the typical $\nu_e$ capture rate
at the freeze-out radius of the $r$-process, $r_{\rm FO}$, with
$\lambda_{\nu_e}\propto r^{-2}$. In derving Eq. (9), we have made the
approximation that the temperature decreases as $r^{-1}$ in the wind
with a constant dynamic scale $\tau_{\rm dyn}\sim r/v$.

However, previous studies of the $r$-process also indicate that the
nuclear flow is dominantly controlled by the $\beta$-decays at the
freeze-out of the $r$-process \cite{kratz}. Therefore, we also
have 
\begin{equation}
\lambda_{\nu_e}(r_{\rm FO})<\lambda_\beta,
\end{equation}
where $\lambda_\beta\sim 3$ s$^{-1}$ is the typical $\beta$-decay rate
of the waiting-point nuclei (see Table 4 in Ref. \cite{full}).
In fact, Eq. (10) was used to constrain the location of the $r$-process
\cite{full}. For typical $\nu_e$ luminosities and spectra, Eq. (10)
gives a lower limit on $r_{\rm FO}$ of $\sim100$ km. Combining Eqs. (9)
and (10), we find that $\nu_e$ captures can speed up the $r$-process
without affecting its freeze-out if
\begin{equation}
\tau_{\rm dyn}>{1\over4\lambda_\beta}.
\end{equation}
While the quantitative effects of $\nu_e$ captures during the $r$-process
can be assessed only in a detailed numerical calculation, some
implications of Eq. (11) are readily available. First of all,
Eq. (11) tells us that in a wind with a constant dynamic time scale
as assumed in deriving Eq. (9), a good $r$-process is possible
only if $\tau_{\rm dyn}>0.1$ s, for which there is enough time
to complete the $r$-process with the help from $\nu_e$ captures.
In this case, a low $Y_e$ may be the last hope for the $r$-process
in the neutrino-driven wind. On the other hand, 
if $\tau_{\rm dyn}<0.1$ s is needed to avoid the undesirable effects of 
$\alpha$-particles on $Y_e$ and the neutron-to-seed ratio discussed
in Sec. 4, the wind has to slow down considerably between the 
$\alpha$-process and the $r$-process. This may be accomplished by
imposing an outer boundary temperature of $T_b\sim 10^9$ K on the
wind \cite{qiawoo}. From $\dot M=4\pi r^2\rho v$ and $S\propto T^3/\rho$,
the velocity decreases according to $v\propto r^{-2}$ after the
temperature in the wind reaches $T_b$ for the first time at radius $r_1$.
The time available for the $r$-process is approximately given by
\begin{equation}
\Delta t\sim\int_{r_1}^{r_{\rm FO}}{dr\over v}\sim
{\tau_{\rm dyn}\over 3}\left[\left({r_{\rm FO}\over r_1}\right)^3-1\right],
\end{equation}
where we have assumed $r_1/v_1\sim \tau_{\rm dyn}$, with
$v_1$ the velocity at $r_1$.
Without significant $\nu_e$ captures, the standard picture
requires $\Delta t\sim 1\ {\rm s}>\lambda_\beta^{-1}$ 
to complete the $r$-process. For $\tau_{\rm dyn}<0.1$ s, the freeze-out
radius of the $r$-process has to be large to give $\Delta t\sim 1$ s.

Finally, we also mention the effects of the supernova neutrino flux
after the $r$-process freezes out. The post-processing of the $r$-process
nuclei by supernova neutrinos was first suggested in Ref. \cite{dom}.
Here we briefly discuss a somewhat different kind of post-processing,
which takes place before the progenitor nuclei successively $\beta$-decay
to the $r$-process nuclei observed in nature. We note that the progenitor
nuclei are left in highly-excited states after $\nu_e$ captures. Being
neutron-rich, these nuclei de-excite by emitting several neutrons
\cite{qhlv}. Neutral-current excitations by $\nu_{\mu(\tau)}$ and 
$\bar\nu_{\mu(\tau)}$ can also induce neutron emissions from the
progenitor nuclei \cite{qhlv}. With adequate exposure to the neutrino
flux, these neutrino-induced neutron emissions may modify the $r$-process
abundance distribution. Detailed discussions of the possible modifications
will be presented elsewhere \cite{qhlv}.

\section{CONCLUSIONS}

We have discussed various neutrino-nucleus interactions in connection
with the $r$-process in the neutrino-driven wind. We find that a
short dynamic time scale in the wind, which is needed to avoid the
undesirable effects of $\alpha$-particles on $Y_e$ and the 
neutron-to-seed ratio, has to be accompanied by some mechanism to
contain the wind during the $r$-process. 
We also find that a low $Y_e$, which counteracts the undesirable
effects of $\alpha$-particles, possibly
arises when the neutron star is almost fully deleptonized.
We conclude that a low $Y_e$ and/or a short dynamic time scale
may be required to give a successful $r$-process in the neutrino-driven
wind.

\end{document}